\documentclass[sigconf]{acmart}
\AtBeginDocument{%
  }

\usepackage{multirow}
\usepackage{tabularx}

\setcopyright{acmlicensed}
\copyrightyear{2025}
\acmYear{2025}
\acmDOI{10.1145/3731715.3733495}
\acmConference[ICMR'25]{The 15th ACM International Conference on Multimedia Retrieval}{June 30--July 03, 2025} {Chicago, USA}
\acmISBN{979-8-4007-1877-9/25/06}

\begin{document}

%%
%% The "title" command has an optional parameter,
%% allowing the author to define a "short title" to be used in page headers.
\title{TeMTG: Text-Enhanced Multi-Hop Temporal Graph Modeling for Audio-Visual Video Parsing}

%%
%% The "author" command and its associated commands are used to define
%% the authors and their affiliations.
%% Of note is the shared affiliation of the first two authors, and the
%% "authornote" and "authornotemark" commands
%% used to denote shared contribution to the research.
% \author{Yaru Chen}
% % \authornote{Both authors contributed equally to this research.}
% \email{yaru.chen@surrey.ac.uk}
% \orcid{0009-0001-6648-0399}
% \affiliation{%
%   \institution{University of Surrey}
%   \city{Surrey}
%   \state{United Kingdom}
%   % \country{UK}
% }
\author{Yaru Chen}
% \authornote{Both authors contributed equally to this research.}
\email{Aria_yc@126.com}
\orcid{0009-0001-6648-0399}
% \author{G.K.M. Tobin}
% \authornotemark[1]
% \email{webmaster@marysville-ohio.com}
\affiliation{%
  \institution{University of Surrey}
  \city{Surrey}
  % \state{United Kingdom}
  \country{United Kingdom}
}

\author{Peiliang Zhang}
% \authornote{Both authors contributed equally to this research.}
\email{cheungbl@ieee.org}
\orcid{0000-0002-1826-3349}
% \author{G.K.M. Tobin}
% \authornotemark[1]
% \email{webmaster@marysville-ohio.com}
\affiliation{%
  \institution{Wuhan University of Technology}
  \city{Wuhan}
  % \state{Hubei}
  \country{China}
}

\author{Fei Li}
% \authornote{Both authors contributed equally to this research.}
\email{leefly072@126.com}
\orcid{0000-0002-8440-359X}
% \author{G.K.M. Tobin}
% \authornotemark[1]
% \email{webmaster@marysville-ohio.com}
\affiliation{%
  \institution{University of Wisconsin-Madison}
  \city{Madison}
  % \state{Hubei}
  \country{United States}
}

\author{Faegheh Sardari}
% \authornote{Both authors contributed equally to this research.}
\email{f.sardari@surrey.ac.uk}
\orcid{0000-0002-9134-0427}
% \author{G.K.M. Tobin}
% \authornotemark[1]
% \email{webmaster@marysville-ohio.com}
\affiliation{%
  \institution{University of Surrey}
  \city{Surrey}
  % \state{Hubei}
  \country{United Kingdom}
}

\author{Ruohao Guo}
% \authornote{Both authors contributed equally to this research.}
\email{ruohguo@stu.pku.edu.cn}
\orcid{0000-0002-1091-272X}
% \author{G.K.M. Tobin}
% \authornotemark[1]
% \email{webmaster@marysville-ohio.com}
\affiliation{%
  \institution{Peking University}
  \city{Beijing}
  % \state{Hubei}
  \country{China}
}

\author{Zhenbo Li}
% \authornote{Both authors contributed equally to this research.}
\email{lizb@cau.edu.cn}
\orcid{0000-0003-2914-1192}
% \author{G.K.M. Tobin}
% \authornotemark[1]
% \email{webmaster@marysville-ohio.com}
\affiliation{%
  \institution{China Agricultural University}
  \city{Beijing}
  % \state{Hubei}
  \country{China}
}

\author{Wenwu Wang}
% \authornote{Both authors contributed equally to this research.}
\email{w.wang@surrey.ac.uk}
\orcid{0000-0002-8393-5703}
% \author{G.K.M. Tobin}
% \authornotemark[1]
% \email{webmaster@marysville-ohio.com}
\affiliation{%
  \institution{University of Surrey}
  \city{Surrey}
  % \state{Hubei}
  \country{United Kingdom}
}

%%
%% By default, the full list of authors will be used in the page
%% headers. Often, this list is too long, and will overlap
%% other information printed in the page headers. This command allows
%% the author to define a more concise list
%% of authors' names for this purpose.
\renewcommand{\shortauthors}{Yaru Chen et al.}

%%
%% The abstract is a short summary of the work to be presented in the
%% article.
\begin{abstract}
  Audio-Visual Video Parsing (AVVP) task aims to parse the event categories and occurrence times from audio and visual modalities in a given video. 
% The Audio-Visual Video Parsing (AVVP) task aims to identify event categories and their occurrence times by leveraging both audio and visual modalities in a given video.
Existing methods usually focus on implicitly modeling audio and visual features through weak labels, without mining semantic relationships for different modalities and explicit modeling of event temporal dependencies. 
% Existing approaches primarily focus on implicitly modeling audio-visual features using weak labels but fail to explore semantic relationships between modalities or explicitly model temporal dependencies of events. 
This makes it difficult for the model to accurately parse event information for each segment under weak supervision, especially when high similarity between segmental modal features leads to ambiguous event boundaries. 
% This limitation makes it challenging for models to accurately parse event information for each video segment under weak supervision, particularly when high feature similarity across segments results in ambiguous event boundaries.
Hence, we propose a multimodal optimization framework, TeMTG, that combines text enhancement and multi-hop temporal graph modeling.
% To address these challenges, we propose TeMTG, a multimodal optimization framework that integrates text enhancement and multi-hop temporal graph modeling. 
Specifically, we leverage pre-trained multimodal models to generate modality-specific text embeddings, and fuse them with audio-visual features to enhance the semantic representation of these features. 
% Specifically, we employ pre-trained multimodal models to generate modality-specific text embeddings, which are then fused with audio-visual features to enhance their semantic representations. 
In addition, we introduce a multi-hop temporal graph neural network, which explicitly models the local temporal relationships between segments, capturing the temporal continuity of both short-term and long-range events. 
% Furthermore, we introduce a multi-hop temporal graph neural network, which explicitly captures local temporal relationships between segments, preserving both short- and long-range event continuity.
Experimental results demonstrate that our proposed method achieves state-of-the-art (SOTA) performance in multiple key indicators in the LLP dataset.
\end{abstract}

%%
%% The code below is generated by the tool at http://dl.acm.org/ccs.cfm.
%% Please copy and paste the code instead of the example below.
%%
% \ccsdesc[500]{Do Not Use This Code~Generate the Correct Terms for Your Paper}
% \ccsdesc[300]{Do Not Use This Code~Generate the Correct Terms for Your Paper}
% \ccsdesc{Do Not Use This Code~Generate the Correct Terms for Your Paper}
% \ccsdesc[100]{Do Not Use This Code~Generate the Correct Terms for Your Paper}

%%
%% Keywords. The author(s) should pick words that accurately describe
%% the work being presented. Separate the keywords with commas.
%% A "teaser" image appears between the author and affiliation
%% information and the body of the document, and typically spans the
%% page.
% \begin{teaserfigure}
%   \includegraphics[width=\textwidth]{sampleteaser}
%   \caption{Seattle Mariners at Spring Training, 2010.}
%   \Description{Enjoying the baseball game from the third-base
%   seats. Ichiro Suzuki preparing to bat.}
%   \label{fig:teaser}
% \end{teaserfigure}

% \received{20 February 2007}
% \received[revised]{12 March 2009}
% \received[accepted]{5 June 2009}
%%
%% This command processes the author and affiliation and title
%% information and builds the first part of the formatted document.
\maketitle
\vspace{-5mm}
\section{Introduction}
\begin{figure}[t] 
\includegraphics[width=0.38\textwidth]{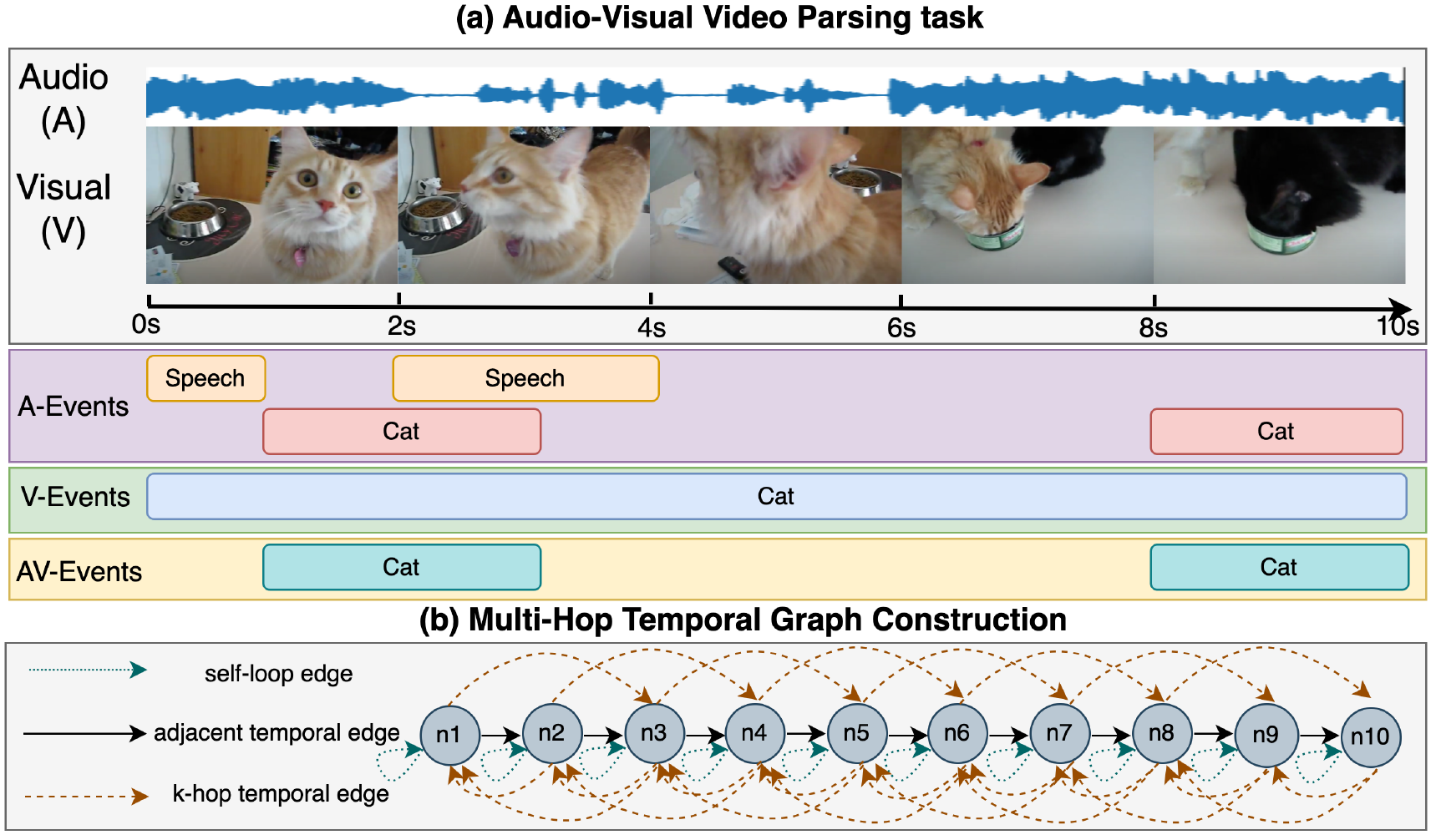} % 调整图片宽度
\vspace{-3mm}
    \caption{(a)~Illustration of the AVVP task. (b)~Structure for Multi-Hop Temporal Graph~(assume $K=2$).} 
    \label{fig:motivation}
    \vspace{-7mm}
\end{figure}

In Audio-Visual Video Parsing~(AVVP)~\cite{tian2020unified} task, our goal is not only to detect what events occur at what times, but also to determine which modality detects the event. Compared with other related tasks~\cite{zeng2024anchor,li2024boosting,guo2024open,li2022drcnet}, a distinguishing characteristic of this task is the temporal asynchrony between events that occur in different modalities. As shown in Fig.~\ref{fig:motivation}~(a), we see cats for 10 seconds, while we hear it between 1-3 and 8-10 s, and we can still hear the sound of speech even if no one appears in the video. Therefore, events are often categorized into three types: audio events, visual events, and audio-visual events. AVVP is also trained under the weakly supervised scenario, which means that the model is trained using only limited video-level event labels. This setting increases the difficulty for models to learn the temporal details and modality correlations. AVVP can benefit a variety of real-world applications, such as intelligent video surveillance, content-based video indexing.

A baseline method~\cite{tian2020unified} was developed for the AVVP task by employing hybrid attention networks~(HAN), where multi-modal multiple instance learning~(MMIL) is used to aggregate the multi-modal temporal contexts, together with the identification and suppression of noisy labels for each modality. Subsequently, the researchers~\cite{gao2023collecting,sardari2024coleaf} explored contrastive learning, distillation learning, and other techniques to connect semantically similar segments within and between modalities. With the emergence of large-scale pre-trained models, Lai et al.~\cite{lai2023modality} utilized pre-trained CLAP~\cite{wu2023large} and CLIP~\cite{radford2021learning} to extract features and generate segment-level pseudo-labels. Other researchers have explored the use of pseudo labels as references for the model to distinguish the semantics of each event that occurred in each segment~\cite{zhou2024label,fan2024revisit}.

% Mo et al.~\cite{mo2022multi} aggregated unimodal features through an explicit semantic grouping strategy and used cross-modal grouping for modality-aware prediction. Faegheh et al.~\cite{sardari2024coleaf} proposed a method based on contrastive learning and collaborative optimization of dual branches, using the reference branch to guide the anchor branch to learn the relationship between audio and video events. Following this, Zhou et al.~\cite{zhou2024label} projects audio-visual segments into an independent semantic label embedding space and incorporate cross-modal interactions along with an EIoU-based semantic similarity loss to enhance the model's decoding discriminability and interpretability. 

% Previous studies have focused on  semantic relationships between video segments without explicitly modeling the temporal dependencies between segments~\cite{gao2023collecting,chen2024cm}.
Previous studies have explored the connections between the categories of events~\cite{mo2022multi,chen2024cm}, but not the propagation and continuation of events in the temporal dimension. In addition, large-scale pre-trained models have been used to provide semantic information~\cite{zhou2024advancing,jiang2024resisting}, which is used as pseudo labels and classification auxiliary. However, they have not been integrated deeply into feature representation. As a result, semantic consistency within the features cannot be guaranteed, causing potential misalignment in audio-visual feature fusion. Recently, text embeddings have been used to improve multimodal representation learning~\cite{wang2024link}. This method focuses primarily on encoding event-related semantics while neglecting background information, which, however, may contain crucial contextual cues to distinguish events. 

% To address the above issues, we propose TeMTG, which combines text-enhanced semantic guidance and multi-hop temporal graph modeling to improve the performance of weakly supervised AVVP. First, we introduce a feature fusion mechanism based on text enhancement, leveraging large-scale pre-trained models to obtain segment-level pseudo labels, thus generating text embeddings for different modalities. These embeddings are then fused through a modality-specific multi-layer perceptron (MLP)~\cite{taud2017multilayer}, providing explicit semantic guidance for audio and visual features and improving the discriminability of multimodal representations. Second, we construct a multi-hop (K-hop) temporal graph, explicitly modeling segment-wise temporal dependencies. By establishing K-hop temporal edges, our model can capture audio-visual event correlations from temporal sequences, thereby enhancing its ability in temporal reasoning. Experimental results demonstrate that our method effectively addresses the limitations of existing AVVP approaches and achieves SOTA performance.
To tackle these challenges, we propose TeMTG, which combines text-enhanced semantic guidance with multi-hop temporal graph modeling for weakly supervised AVVP. We first introduce a fusion mechanism that leverages large-scale pre-trained models to generate segment-level pseudo labels and corresponding text embeddings, which are refined by a modality-specific multi-layer perceptron~(MLP)~\cite{taud2017multilayer} to enhance semantic guidance and feature discriminability. Then, we construct a K-hop temporal graph to explicitly model segment-wise dependencies. By linking segments through K-hop edges, our model captures audio-visual correlations across time, improving temporal reasoning. Experimental results show that TeMTG effectively addresses AVVP limitations and achieves SOTA performance.

 \vspace{-3.5mm}
\section{Proposed Methodology}
%\subsection{Preliminary}
%\vspace{-1mm}
In the AVVP task, a video clip \textit{S} can be divided into $T$ segments, represented as \(S=\{A_{t}, V_{t}\}_{t=1}^T\), where $A_t$ and $V_t$ denote the audio and visual features of the $t$-th segment. The task requires segmenting events into three categories: audio events $y_t^a \in \lbrace 0,1 \rbrace^C$, visual events $y_t^v \in \lbrace 0,1 \rbrace^C$, and audio-visual events $y_t^{av} \in \lbrace 0,1 \rbrace^C$, where $C$ is the total number of event classes. An event is considered audio-visual if it appears in both modalities simultaneously, i.e., \(y_t^{av} = y_t^a * y_t^v\). During training, only weak video-level labels $y \in \lbrace 0,1 \rbrace^C$ are provided. The goal of AVVP is, given a video clip divided into \textit{T} segments of audio‐visual features, to determine for each segment whether an event (among \textit{C} possible classes) occurs in the audio modality, the visual modality, or both.
\begin{figure*}[h]
    \centering    \includegraphics[width=0.85\textwidth]{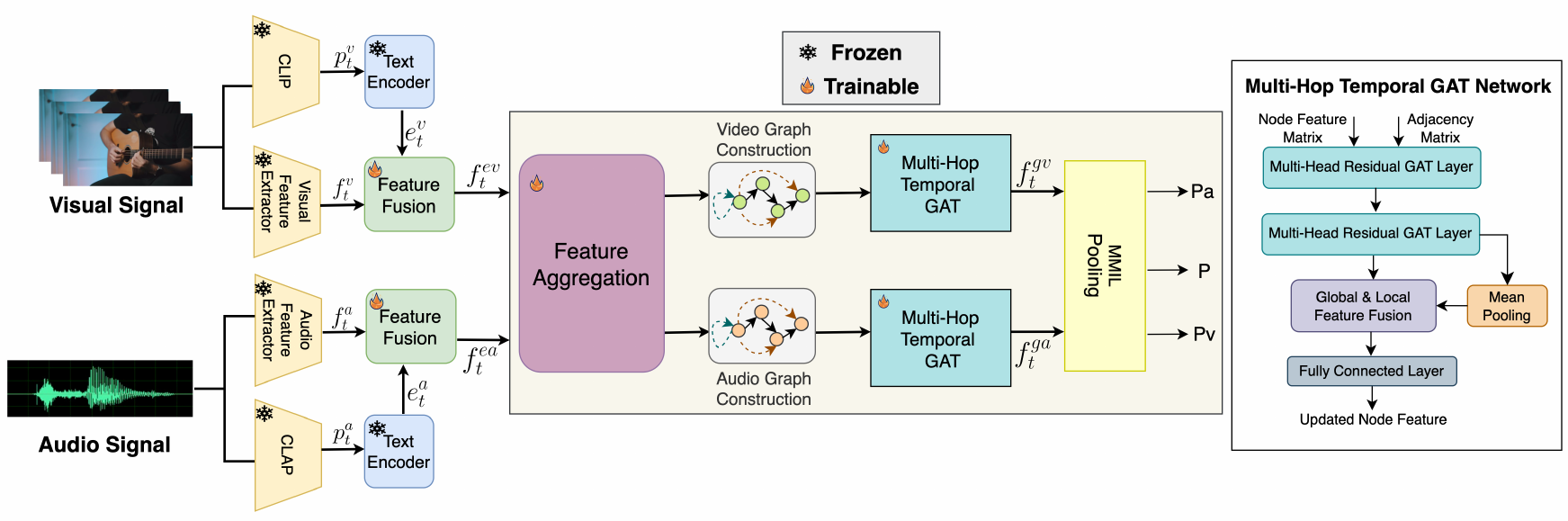} % 调整图片宽度
     % \caption{The architecture of TeMTG. The audio and visual features are fused with text embeddings and processed through the Feature Aggregation Module. Then, multi-hop temporal graphs with Multi-Head Residual GAT layers are used to model the temporal dependencies of the events.}
     \vspace{-3mm}
     \caption{TeMTG architecture: Audio and visual features are fused with text embeddings, aggregated, and then processed by multi-hop temporal graphs with multi-head residual GAT layers to model event dependencies.
}
    \label{fig:model}
    \vspace{-5mm}
\end{figure*}
 \vspace{-3mm}
\subsection{Framework}
We adopt CoLeaF~\cite{sardari2024coleaf} as our baseline model and propose a novel multimodal optimization framework that integrates text enhancement and multi-hop temporal graph modeling. As shown in Fig.~\ref{fig:model}, we first use the text encoder to generate text embeddings for the audio and visual stream of each video segment, respectively. These embeddings are then fused with the audio and visual features, respectively, through a feature fusion module. Next, the fused multimodal features are fed into the feature aggregation module which adopts self-attention and cross-attention to preserve unimodal feature learning and enhance cross-modal interactions. Then we construct a multi-hop temporal graph and propagate information using multi-head graph attention (GAT)~\cite{velivckovic2018graph} to model both short-term continuity and long-term dependencies among video segments. Finally, we employ MMIL pooling~\cite{tian2020unified} to aggregate temporal features and generate the final video-level predictions, including audio predictions $P_a$, visual predictions $P_v$, and joint audio-visual predictions $P$. As CoLeaF used two branches for feature aggregation, we placed our proposed multi-hop temporal GAT after each branch.
 \vspace{-3mm}
\subsection{Text-Enhanced Multimodal Feature Fusion}
To enhance the semantic representation of the audio-visual features, we introduce text embeddings to provide explicit semantic guidance, effectively mitigating the limitations of weakly supervised learning. %and improving the model’s generalization ability by incorporating external knowledge.
Specifically, 
% for the input video \textit{S}, 
inspired by~\cite{lai2023modality}, we first use CLAP and CLIP to generate pseudo labels at segment level $p_t^a, p_t^v \in \lbrace 0,1 \rbrace^C$. Then we convert each pseudo label into a text description in the following format: "\textit{This is the sound of} x \textit{ audio event}" or "\textit{This is the image of} x \textit{ visual event}". If a segment contains multiple audible or visible events, we concatenate their corresponding descriptions with conjunctions (e.g.~"\textit{This is the sound of event A and event B}"). If a segment contains no events, the corresponding text is set as "\textit{There is no sound in the segment}" or "\textit{There is no event in the image}". Then, we feed these texts into the text branches of CLAP and CLIP to generate the corresponding text embeddings $e_t^a,e_t^v\in \mathbb{R}^{b\times T\times d}$, in which~\textit{b} is the batch size, and~\textit{d} is the dimension. %These embeddings are used to enhance the semantics of the audio and visual modalities.

Afterwards, we design a modality-specific fusion strategy based on MLP~\cite{taud2017multilayer} to effectively integrate textual semantic information into audio-visual feature representations. We first get the audio and visual feature representations $f_t^a, f_t^v\in \mathbb{R}^{b\times T\times d}$ from their feature extractors, and then concatenate the features and their text embeddings along the feature dimension to obtain the fused input:
\begin{equation}
\small
    z_t^a = ({f_t^a}\parallel e_t^a)\in \mathbb{R}^{b\times T\times 2d}
\end{equation}
where $\parallel$ is the concatenate operation. This concatenated feature is then mapped through a two-layer MLP to obtain the fused audio feature:
\begin{equation}
\small
    f_t^{ea}= \sigma \left ( W_2\left ( ReLU\left ( W_1z_t^a+b_1 \right ) \right ) +b_2\right )
\end{equation}
where $W_1\in \mathbb{R}^{2d\times m}$ and $W_2\in \mathbb{R}^{d\times m}$, are the weight matrices of the two-layer MLP, $b_1$ and $b_2$ are bias terms, $m$ is the hidden layer dimension, and $\sigma\left ( \cdot  \right )$ denotes the LayerNorm operation. Similarly, the fusion process for the visual modality is defined as follows:
\begin{equation}
\small
    z_t^v = ({f_t^v}\parallel e_t^v)\in \mathbb{R}^{bs\times T\times 2d} 
\end{equation}
\begin{equation}
\small
    f_t^{ev}= \sigma \left ( W_2\left ( ReLU\left ( W_1z_t^a+b_1 \right ) \right ) +b_2\right )
\end{equation}
 Finally, we apply a linear layer to project the fused audio and visual features back to the original feature dimension $d$, ensuring compatibility with the input of the downstream task and maintaining consistency with the original feature space after fusion. 
  \vspace{-3mm}
 \subsection{Multi-Hop Temporal Graph Modeling}
 % To capture both short‐term continuity and long‐range dependencies among video segments, we propose a Multi‐Hop Temporal Graph~(MTG) modeling approach. 
 % % Rather than rely solely on self-attention mechanisms, 
 % MTG explicitly constructs a structured graph in which each segment is linked not only to its immediate neighbors, but also to multiple preceding/following segments within a predefined range~$K$. As shown in Fig.~\ref{fig:motivation}~(b), $K$ is a hyperparameter that controls the maximum temporal connection range, ensuring information propagation across multiple time steps. Note that the value of $K$ for the audio and visual graphs can be different, allowing flexible temporal modeling in each modality. This bidirectional design allows each node to aggregate contextual information from both the previous and next segments, thus enhancing temporal relation reasoning. 
 To model both short- and long-term dependencies, we propose a Multi-Hop Temporal Graph (MTG), which links each segment not only to its neighbors but also to others within a temporal range~$K$. This bidirectional, modality-specific design enables flexible and effective temporal modeling by allowing each node to aggregate contextual information from both past and future segments, enhancing temporal relation reasoning.

 % To explicitly model both short-term continuity and long-range dependencies between video segments, we propose a Multi-Hop Temporal Graph Modeling approach. Unlike existing methods that rely on self-attention mechanisms for temporal modeling, we construct a structured temporal graph, where each segment is connected to its adjacent segments and multiple preceding/following segments within a predefined range \textit{K}. This design enables information propagation across both local and global temporal contexts, ensuring a more comprehensive temporal dependency modeling.

Specifically, for the input features $f_t^{ea}$ and $f_t^{ev}$, we construct the temporal graphs $G^a=\left ( {N^a},{E^a} \right)$ and $G^v=\left ( {N^v},{E^v} \right)$ for each video at the segment level, where the nodes $ N^a=\left\{ n_1^a,n_2^a,\cdots ,n_T^a\right\},{n_t^a}\in~\mathbb{R}^d$ and $N^v=\left\{ n_1^v,n_2^v,\cdots ,n_T^v\right\},{n_t^v}\in\mathbb{R}^d$ represent the audio and visual features representations of each segment, and edges ${E^a}$ and ${E^v}$ are the temporal relationships between the segments. Hence, for each audio or visual node $n_t$, we first define their K-Hop bidirectional temporal connection edges as follows:
% \begin{equation}
% \small
%     E_t^a,E_t^v= \left\{ \left ( n_t,n_{t-k} \right ),\left ( n_t,n_{t+k} \right )\mid 0<k\leqslant K, k\leqslant t<seg-k\right\}
% \end{equation}
\begin{equation}
\begin{aligned}
    E^a = &\bigl\{\,(n_t, n_{t-k}) \;\big|\; 1 \leq t \leq T,\; 1 \leq k \leq K,\; t - k \geq 1\bigr\} \\
    &\cup \bigl\{\,(n_t, n_{t+k}) \;\big|\; 1 \leq t \leq T,\; 1 \leq k \leq K,\; t + k \leq T\bigr\}.
\end{aligned}
\end{equation}
where $k$ is the hop distance. Additionally, each node has a self-loop to preserve its original information. The final adjacency matrix for audio and visual temporal graph $A^a$ and $A^v$ are:
\begin{equation}
\small
    A_{ij}^a, A_{ij}^v=\left\{\begin{matrix}
        1, & if\space\space\space\space 0\leq{j-i}\leq K \\
        0, & otherwise \\ 
        \end{matrix}\right.
\end{equation}
where $i$ and $j$ are the index of the nodes.

Then we employ the multi-head residual GAT to perform feature aggregation on the constructed temporal graph, allowing nodes to dynamically weight different time steps based on contextual information. Specifically, given the multi-hop temporal graph $G$ and its adjacency matrix $A$, the attention weight between two connected nodes $n_i$ and $n_j$ is computed as follows:
\begin{equation}
\small
\alpha _{ij}=\frac{exp\left ( LeakyReLU\left ( a^{(h)T}\left [ W^{(h)}{n_i}\parallel W^{(h)}{n_j} \right ] \right ) \right )}{\sum_{k\in \phi _i}exp\left ( LeakyReLU\left ( a^{(h)T}\left [ W^{(h)}{n_i}\parallel W^{(h)}{n_k} \right ] \right ) \right )} 
\end{equation}
where $\cdot^T$ represents transposition, $h$ is the multi-head attention, $W\in \mathbb{R}^{d\times d}$ is a learnable weight matrix, $a\in\mathbb{R}^{2d\times 1}$ is the attention vector, and $\phi_i$ is the set of neighbors of the node $i$. The final update of the node features is as follows:
\begin{equation}
\small
 n'_i=\varepsilon \left ( \frac{1}{H}\sum \nolimits_{h=1}^{H}\sum \nolimits_{j\in \phi _i}\alpha^{(h)}_{ij} \cdot {W^{(h)}{n_j}} \right )
\end{equation}
where $\varepsilon(\cdot)$ is a nonlinear activation function, and $H$ is number of heads for multi-head attention. To enhance model stability, we incorporate residual connections, batch normalization, and dropout in the design of K-hop GAT layer, further optimizing the effectiveness of temporal information propagation.

After GAT propagation, we use global mean pooling to extract the temporal representation of the entire video, enhancing cross-node information integration. Subsequently, we use an MLP to further fuse the global video features with each feature node, enabling each node to model local temporal relationships while also perceiving the global context of the entire video, which can effectively model both short-term and long-term event dependencies. %This design bridges local continuity with global interactions, enabling information to propagate across multiple temporal scales and providing a more comprehensive modeling of dependencies throughout the video.

\begin{table*}[t]
  \caption{The performance of TeMTG and comparative methods in AVVP, with the best results highlighted in \textbf{bold} and the second results highlighted in \underline{text}.}
  \vspace{-3mm}
  \label{tab:main-results}
  \renewcommand{\arraystretch}{0.8}
  \begin{tabular}{l|c|ccccc|ccccc}
    \toprule
    \multirow{2}{*}{Model} & \multirow{2}{*}{Venue} & \multicolumn{5}{c|}{Segment-level (\%)} & \multicolumn{5}{c}{Event-level (\%)}\\
    \cmidrule{3-7} \cmidrule{8-12}
    &  & A & V & AV & Type@AV & Event@AV & A & V & AV & Type@AV & Event@AV \\
    \midrule
    HAN~\cite{tian2020unified} & ECCV'20 & 60.1 & 52.9 & 48.9 & 54.0 & 55.4 & 51.3 & 48.9 & 43.0 & 47.7 & 48.0 \\
    % \cmidrule{2-19}
    MGN~\cite{mo2022multi} & NeurIPS'22 & 60.8 & 55.4 & 50.0 & 55.1 & 57.6 & 52.7 & 51.8 & 44.4 & 49.9 & 50.0 \\
    % \cmidrule{2-19}
    MA~\cite{wu2021exploring} & CVPR'21 & 60.3 & 60.0 & 55.1 & 58.9 & 57.9 & 53.6 & 56.4 & 49.0 & 53.0 & 50.6 \\
    % \cmidrule{2-19}
    CMPAE~\cite{gao2023collecting} & CVPR'23 & 64.2 & 66.2 & 59.2 & 63.3 & 62.8 & 56.6 & 63.7 & 51.8 & 57.4 & 55.7 \\
    % \cmidrule{2-19}
    CoLeaF~\cite{sardari2024coleaf} & ECCV'24 & 64.2 & 67.1 & 59.8 & 63.8 & 61.9 & 57.1 & 64.8 & 52.8 & 58.2 & 55.5 \\
    % \cmidrule{2-19}
    LEAP~\cite{zhou2024label} & ECCV'24 & 64.8 & 67.7 & 61.8 & 64.8 & 63.6 & 59.2 & 64.9 & \underline{56.5} & 60.2 & 57.4 \\
    % \cmidrule{2-19}
    VALOR++~\cite{lai2023modality} & NeurIPS'23 & 68.1 & 68.4 & 61.9 & 66.2 & 66.8 & 61.2 & 64.7 & 55.5 & 60.4 & 59.0 \\
    % \cmidrule{2-19}
    LSLD+~\cite{fan2024revisit} & NuerIPS'23 & 68.7 & \underline{71.3} & \underline{63.4} & 67.8 & 68.2 & 61.5 & \underline{67.4} & 55.9 & \underline{61.6} & 60.6\\
    % \cmidrule{2-19}
    NREP~\cite{jiang2024resisting} & TNNLS'24 & \underline{70.2} & 70.9 & \textbf{64.4} & \underline{68.5} & \underline{68.8} & \textbf{62.8} & 67.3 & \textbf{57.6} & \textbf{62.6} & \underline{61.1} \\
    \midrule
    TeMTG~(Ours) & - & \textbf{74.4}  & \textbf{72.9}  & 62.0 & \textbf{69.8} & \textbf{74.1} & \underline{61.9} & \textbf{69.0} & 53.2 & 61.4 & \textbf{62.2}  \\
    &  & \textcolor{blue}{(+4.2)} & \textcolor{blue}{(+1.6)} &  & \textcolor{blue}{(+1.3)} & \textcolor{blue}{(+5.3)} &  & \textcolor{blue}{(+1.6)} &  &  & \textcolor{blue}{(+1.1)} \\
  \bottomrule
\end{tabular}
% }
\vspace{-2mm}
\end{table*}
\begin{table}[t]
  \caption{Ablation study for TeMTG. w/o~TE and w/o~MTG mean without TE and MTG respectively.}
  \vspace{-3mm}
  \label{tab:abs-results}
  \setlength{\tabcolsep}{1mm}{
  \renewcommand{\arraystretch}{0.4}
  \begin{tabular}{clccccc}
    \toprule
    % & & \multicolumn{5}{c}{Segment-level (\%)}\\
    % \midrule
    & Method & A & V & AV & Type@AV & Event@AV\\
    \midrule
   \multirow{4}{*}{Segment-level} & CoLeaF$^\dagger$ & 64.2  & 67.4  & 59.9 & 63.8 & 63.3\\
   & \textit{ w/o}~TE & 64.8 & 68.9 & 60.6 & 64.8 & 64.2 \\
   & \textit{ w/o}~MTG & 76.5 & 72.9 & 62.4 & 70.6 & 75.7 \\
    \cmidrule{2-7}
    & TeMTG & 74.4 & 72.9 & 62.0 & 69.8 & 74.1\\
    \midrule
    % & & \multicolumn{5}{c}{Event-level (\%)}\\
    % \midrule
   & Method & A & V & AV & Type@AV & Event@AV\\
    \midrule
   \multirow{4}{*}{Event-level} & CoLeaF$^\dagger$ & 53.2 & 64.1 & 52.4 & 56.6 & 52.7 \\
    & \textit{ w/o}~TE & 53.5 & 65.6 & 52.4 & 57.1 & 53.3\\
    & \textit{ w/o}~MTG & 66.6 & 69.0 & 53.4 & 63.1 & 66.0\\
   \cmidrule{2-7}
   & TeMTG & 61.9 & 69.0 & 53.2 & 61.4 & 62.2\\
  \bottomrule
\end{tabular}
}
\vspace{-8mm}
\end{table}
 \vspace{-3mm}
\section{Experimental Results}
\subsection{Experimental Setup}
\textbf{Dataset and Implementation Details.} We use the LLP dataset~\cite{tian2020unified} to evaluate our framework, which includes 11849 videos with 25 categories and has been widely used in the AVVP task. 
% It sets 10000 videos with weak labels as the training set, 1200 videos as the test set, and 649 videos with fully annotated labels as the validation set. 
Each video is divided into 10 segments and each segment lasts 1~second. We utilize the pre-trained CLAP~\cite{wu2023large} to extract 768-D audio features from the audio signal. We use pre-trained CLIP~\cite{radford2021learning} and 3D ResNet to extract 768-D and 512-D visual features from the visual signal, then fuse the concatenated 2D and 3D visual features. Finally, a linear layer is used to project audio and visual features into the same feature space to facilitate subsequent operations. We set the number of hops $K=4$ for both audio and visual temporal graphs to ensure a balanced temporal dependency modeling across modalities. In addition, 
% we train TeMTG using the Adam optimizer with an initial learning rate of $5\times10^{-4}$ and a batch size of 128, for 10 epochs. The learning rate was decayed by a factor of 0.25 every 6 epochs. 
we performed our experiments using PyTorch on an NVIDIA A100 GPU.\\
% \textbf{Implementation Details.} 
% We utilize the pre-trained CLAP~\cite{wu2023large} to extract 768-D audio features from the audio signal. We use pre-trained CLIP~\cite{radford2021learning} and 3D ResNet to extract 768-D and 512-D visual features from the visual signal, then fuse the concatenated 2D and 3D visual features. Finally, a linear layer is used to project audio and visual features into the same feature space to facilitate subsequent operations. We set the number of hops $K=4$ for both audio and visual temporal graphs to ensure a balanced temporal dependency modeling across modalities. In addition, 
% % we train TeMTG using the Adam optimizer with an initial learning rate of $5\times10^{-4}$ and a batch size of 128, for 10 epochs. The learning rate was decayed by a factor of 0.25 every 6 epochs. 
% we performed our experiments using PyTorch on an NVIDIA A100 GPU.\\
\textbf{Evaluation Metrics} 
Following~\cite{tian2020unified,gao2023collecting}, 
% we use the F1-score to evaluate audio events~(\textbf{A}), visual events~(\textbf{V}) and audio-visual events~(\textbf{AV}), and set mIoU = 0.5 as the threshold. For each type of event, we assess the F1-score at both the segment and event levels. At the segment level, the predicted events are compared with the ground truth for each individual segment. At the event level, a series of consecutive segments corresponding to the same event is considered to be a whole event. Furthermore, $\textbf{Type{@}AV}$ measures the average F1-score across three type of the events, and $\textbf{Event{@}AV}$ computes the F1-score by jointly evaluating all audio and visual events within each video.
we use F1-score to evaluate audio~\textbf{(A)}, visual~\textbf{(V)}, and audio-visual~\textbf{(AV)} events, with mIoU $\geqq $ 0.5 as the threshold. F1-scores are computed at both segment and event levels: the former compares predictions per segment, while the latter considers sequences of segments as complete events. \textbf{Type@AV} averages F1-scores across A, V, and AV, and\textbf{ Event@AV} jointly evaluates all events in a video.
 \vspace{-1mm}
\subsection{Overall Performance Analysis}
Table~\ref{tab:main-results} shows the experimental results of the comparison between our method and the existing SOTA methods in the LLP dataset. From the results, it can be seen that TeMTG has achieved the best performance in multiple key indicators, especially in segment-level parsing, which significantly surpasses existing methods.

% In segment-level evaluation, TeMTG achieved the best performance in parsing audio events (A) and visual events (V), outperforming the previous best method NREP by 4.2\% and 1.6\%. This shows that the text-enhanced feature fusion mechanism effectively improves the representation capability of audio and visual modalities. Moreover, in the Event@AV metric, TeMTG surpassed NREP by 5.3\%, indicating that our multi-hop temporal graph modeling better integrates cross-segment temporal information, thereby enhancing the overall event parsing performance. However, for AV event parsing, TeMTG achieved an F1-score of 62.0\%, which is lower than NREP (64.4\%). This may be because it remains challenging to accurately model the co-occurrence patterns of audio-visual events under weak supervision, despite the feature enhancement with textual information.
In segment-level evaluation, TeMTG achieved the best results for both audio (A) and visual (V) event parsing, outperforming NREP by 4.2\% and 1.6\%, respectively, highlighting the effectiveness of text-enhanced feature fusion. For Event@AV, TeMTG exceeded NREP by 5.3\%, showing that our multi-hop temporal graph better captures cross-segment temporal dependencies. However, for AV event parsing, TeMTG scored 62.0\%, lower than NREP (64.4\%), likely due to challenges in modeling audio-visual co-occurrence under weak supervision, despite textual enhancements.

% In event-level evaluation, TeMTG achieved 69.0\% in visual event (V) parsing, surpassing LSLD+ by 1.6\%, indicating that our method better captures event features in visual modality. However, for audio event (A) parsing, TeMTG obtained an F1-score of 61.9\%, lower than NREP (62.8\%). This decrease may be due to that our temporal graph modeling does not impose additional constraints on the semantic consistency of key audio events, which may lead to residual interference from background noise. For the AV column, our model shows the lower result~(53.2\%) than other methods, that is because the pseudo labels are generated by CLAP and CLIP, which still contain noise and uncertainty in weakly supervised scenarios, especially in the case of AV events, affecting the accuracy of event boundaries. Nevertheless, our method still achieves the best result in the Event@AV metric (62.2\%), demonstrating that although there is room for improvement in individual modality event parsing, TeMTG overall provides a better event parsing capability under weak supervision.
In event-level evaluation, TeMTG achieved 69.0\% in visual event (V) parsing, 1.6\% higher than LSLD+, showing better capture of visual features. However, for audio event~(A) parsing, TeMTG scored 61.9\%, slightly below NREP (62.8\%), likely due to the lack of semantic constraints in our temporal graph model, which may leave residual noise from background sounds. For AV event parsing, TeMTG yielded a lower score (53.2\%) compared to others, as pseudo labels from CLAP and CLIP still carry noise and uncertainty, especially affecting AV boundary accuracy. Nevertheless, TeMTG achieved the highest Event@AV score (62.2\%), demonstrating strong overall parsing ability under weak supervision, despite room for improvement in unimodal performance.
\subsection{Ablation Experiment Analysis}
To verify the effectiveness of the text enhancement mechanism (\textbf{TE}) and multi-hop temporal graph modeling module(\textbf{MTG}) in TeMTG, we removed these two modules for ablation experiments and conducted comparative analysis. Since the experimental results given in the CoLeaF paper are based on the audio and visual features extracted by VGGish and ResNet, to more intuitively compare the performance, we first train CoLeaF using the same input features as TeMTG, namely CoLeaF$^\dagger$. 

As shown in Table~\ref{tab:abs-results}, when only using \textbf{MTG} module at the segment level, compared with CoLeaF$\dagger$, the performance for detecting audio events~(A) has increased from 64.2\% to 64.8\%, visual events (V) increased from 67.4\% to 68.9\%, and Event@AV increased from 63.3\% to 64.2\%. At the event level, audio events (A) increased by 1.3\% and visual events (V) increased by 1.5\%. This indicates that temporal modeling can improve the ability to integrate local temporal information and improve single-modal feature analysis.

% When only set the \textbf{TE} mechanism, the model has a significant improvement in unimodal event parsing, especially for audio events, with an increase of more than 10\% at both segment level and event level. There is also the improvement of 2.5\% in AV event parsing, indicating that text enhancement can also help cross-modal event parsing to a certain extent. In addition, our model has improved by more than 10\% in Event@AV in both segment-level and event-level. Since this indicator focuses on the ability to parse complete events, this shows that by fusing modal features and text embedding, we can not only provide the model with richer event category information, but help the model better learn the consistency of events at different time steps and reduce the number of events that are incorrectly segmented.
With only the TE mechanism enabled, the model shows significant improvement in unimodal event parsing—especially for audio—with over 10\% gains at both segment and event levels. It also achieves a 2.5\% gain in AV event parsing, indicating that text enhancement benefits cross-modal parsing as well. Furthermore, Event@AV improves by over 10\%, suggesting that fusing modal features with text embeddings enriches event category information, helps the model learn temporal consistency, and reduces event segmentation errors.

% In addition, some indicators of TeMTG are slightly lower than those of adding only TE mechanism. This may be because the smoothing effect of temporal modeling weakens the discriminative ability of text enhancement. In the process of local feature aggregation, multi-hop temporal graph modeling may cause the features from some segments to spread to the surrounding segments, resulting in overlapping of the features of some events, thus reducing its discrimination brought by text enhancement. We will study this problem further in future research.
Some TeMTG indicators are slightly lower than using only the TE mechanism, possibly due to temporal modeling's smoothing effect reducing the discriminative power of text enhancement. Multi-hop temporal aggregation may spread features across segments, leading to event overlap and reduced feature distinctiveness. We plan to explore this issue in future work.
 \vspace{-4mm}
\section{Conclusion}
% In this paper, we have presented TeMTG, a multimodal optimization framework that integrates text enhancement and multi-hop temporal graph modeling to improve the weakly supervised AVVP task. The text enhancement mechanism effectively improves the accuracy of event classification between the segments with high similarity, and multihop temporal graph modeling improves the model's temporal reasoning capabilities. Experimental results demonstrate that TeMTG achieves SOTA performance on the LLP dataset, outperforming existing methods in multiple metrics. In addition, we observed the limitations of the proposed method, namely, the smoothing effects in temporal modelling, that requires further study.
We present TeMTG, a multimodal framework combining text enhancement and multi-hop temporal graph modeling for weakly supervised AVVP. Text enhancement improves event classification between similar segments, while temporal modeling enhances reasoning across time. Experiments on the LLP dataset show TeMTG achieves SOTA performance across multiple metrics. However, smoothing effects from temporal modeling remain a limitation, which we plan to explore further.

\vspace{-4mm}
\section*{Acknowledgements}
This work was partly supported by a research scholarship from the China Scholarship Council (CSC). For the purpose of open access, the authors have applied a Creative Commons Attribution (CC BY) license to any Author Accepted Manuscript version arising.

%%
%% The next two lines define the bibliography style to be used, and
%% the bibliography file.
%\bibliographystyle{ACM-Reference-Format}
\bibliographystyle{unsrt}
\balance
\bibliography{sample-base}

\end{document}